\documentstyle[aps,prl,epsfig,amssymb]{revtex}
\begin{document}
\draft
\title{Phase transition in a $2$-dimensional Heisenberg model}
\author{Henk W. J. Bl\"ote~$^{\dag,\ddag}$, Wenan Guo~$^{\S}$
        and Henk J. Hilhorst~$^\ast$} 
\address{$^{\dag}$Faculty of Applied Sciences, Delft University of
Technology,\\  P.O. Box 5046, 2600 GA Delft, The Netherlands}
\address{$^{\ddag}$ Lorentz Institute, Leiden University,
  P.O. Box 9506, 2300 RA Leiden, The Netherlands}
\address{$^{\S}$ Physics Department, Beijing Normal
University, Beijing 100875, P. R. China }
\address{$^{\ast}$ Laboratoire de Physique Th\'eorique,
  B\^atiment 210, Universit\'e de Paris-Sud,\\ 91405 Orsay, France}
\date{\today}
\maketitle
\begin{abstract}
  We investigate the two-dimensional classical Heisenberg model with
  a nonlinear nearest-neighbor interaction 
  $V(\vec{s},\vec{s}\,')=2K[(1+\vec{s}\cdot\vec{s}\,')/2 ]^p$.
  The analogous nonlinear interaction for the XY model was introduced
  by Domany, Schick, and Swendsen, who find that for large $p$ the
  Kosterlitz-Thouless transition is preempted by a first-order transition.
  Here we show that, whereas the standard ($p=1$) Heisenberg model has no
  phase transition, for large enough $p$ a first-order transition appears.
  Both phases have only short range order, but with a correlation
  length that jumps at the transition.
\end{abstract}

The two-dimensional
Heisenberg and XY model are such close relatives that it
has taken a long history of efforts before their properties could be told
apart. Both are special cases, for $n=3$ and $2$, respectively, 
of the O($n$) symmetric Hamiltonian
\begin{equation}
{\mathcal H} = - K \sum_{<i,j>} \vec{s}_{i} \cdot \vec{s}_{j}
\label{Hlin}
\end{equation}
Here $\vec{s}_{i}$ is an $n$-component spin 
of unit length at lattice site $i$, 
the sum is on all pairs of nearest-neighbor sites of a
two-dimensional lattice, and $K=J/k_{\rm B}T$.  
For all $n>1$ the system (\ref{Hlin}) has $d=2$ as
its lower critical dimension.

Bloch's 1930 spin wave argument \cite{Bloch}, put on a firm mathematical
basis only much later by Mermin and Wagner \cite{MW,Mermin}, implies that
neither the XY nor the Heisenberg model can have a 
spontaneously magnetized low-$T$ phase.
The early investigations dealt exclusively
with the Heisenberg model. In 1958 Rushbrooke and Wood, after studying
high-$T$ series \cite{RW}, first remarked that 
in spite of Bloch's argument the
possibility of a phase transition in the Heisenberg model should be taken
seriously. This was reemphasized in 1966 by Stanley and Kaplan
\cite{SK}, who envisage, for the {\it Heisenberg} model, a
low-$T$ phase with an infinite susceptibility. 

In the late 1960's 
the high-$T$ series of the
Heisenberg the XY model were compared \cite{Stanley,Moore}. 
Qualitative similarity was found, but no
general agreement was ever reached about the significance of certain
quantitative differences. A phase transition in either model
continued to be considered by many as only a remote possibility, until
Kosterlitz and Thouless (KT) \cite{KT} demonstrated that there {\it is}
a phase transition in the {\it XY model}
and clarified its topological character.

Since the KT arguments were specific for $n=2$, the 
two-dimensional Heisenberg model
(and, indeed, the Hamiltonian (\ref{Hlin}) for all $n>2$) has from then
on been believed to be without a transition. Further 
support for this view came from the analytical
low-$T$ renormalization group approach developed by Polyakov
\cite{Polyakov}, Br\'ezin and Zinn-Justin \cite{BZJ}, and Nelson and
Pelcovitz \cite{NP}, and from Monte Carlo renormalization 
due to Shenker and Tobochnik \cite{ST}. The absence of a rigorous
proof has however left room for arguments
(\cite{PS} and references therein)
that the Heisenberg model (Eq.\,(\ref{Hlin}) with $n=3$)
may after all have a phase transition; this is not, however,
our point of view.\\

Here we consider the O($3$) symmetric Hamiltonian
\begin{equation}
{\mathcal H} =-\sum_{<i,j>} V(\vec{s}_{i}\cdot \vec{s}_{j})
\label{ham1}
\end{equation}
where $V$ is an arbitrary nonlinear function. 
For reasonable choices of $V$ (in a sense not {\it a priori}
clear) one expects that (\ref{ham1}) is
in the same universality class as
the standard ``linear'' O($n$) model (\ref{Hlin}). 
Expression (\ref{ham1}) is interesting for at least two reasons.

First, the freedom to choose $V$ is a key ingredient in theoretical
analyses by Villain \cite{V} of the O(2) model and by Domany {\it
et al.} \cite{DMNS} and Nienhuis \cite{N} of the O($n$) loop model.
For $n>2$ the latter model does undergo a phase transition \cite{GBW}
which corresponds to a hard-hexagon-like ordering of the loops. But in
spin language the transition
appears to occur in an unphysical parameter
region with negative Boltzmann weights. It does not provide evidence
for a phase transition in O($n$) spin models with $n>2$.

The second reason of interest in (\ref{ham1}) comes from the
relevance of the KT theory for the melting of thin adsorbed layers.
The difficulty encountered in observing the predicted \cite{NH} 
hexatic phase, whether
experimentally or in simulations, was suspected by some to be due to
the KT transition being preempted by a first order transition as a
consequence of various nonlinearities not
incorporated in the theory. Domany et al. (DSS) \cite{DSS} therefore
investigated an O(2) symmetric XY model with a specific nonlinearity
controlled by a parameter $p$, {\it viz.} 
\begin{equation}
V(\vec{s}_{i} \cdot \vec{s}_{j})=2K[(1+\vec{s}_{i}\cdot \vec{s}_{j})/2 ]^p
\label{hdef}
\end{equation}
(our $p$ is their $p^2$).
Indeed DSS found by Monte Carlo simulations that for strong enough
nonlinearity $(p\sim 50)$ the KT transition is replaced with a
first-order one from the massless low-$T$ phase to a high-$T$ phase
with exponentially decaying correlations. While this suggests that
melting via a hexatic phase may similarly be preempted by a first-order
transition, the DSS result has been subject to controversy \cite{JMN}.\\
 
Here we confront again the XY and Heisenberg model.
We have Monte Carlo simulated the latter with the nonlinear interaction
(\ref{hdef}) on square $L \times L$ periodic lattices.
Randomly chosen orientations 
are accepted with Metropolis-type probabilities. 
Slow relaxation at low $T$ limits the largest system
size to about $L=200$.

\begin{figure}
\epsfig{figure=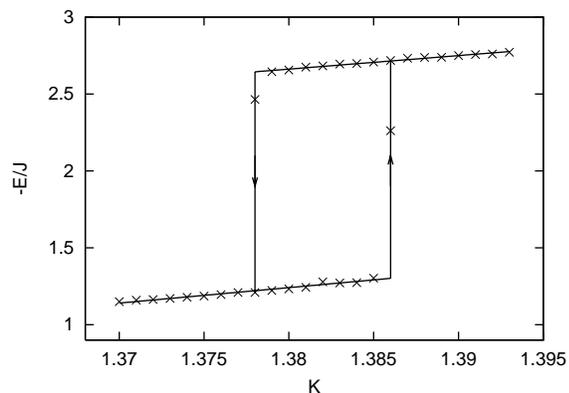}
\caption[xxx]
{Energy per spin versus coupling for a periodic system of size $L=48$ and
$p=20$. The energy discontinuity and the hysteresis indicate a
first-order phase transition. Each data point results from
$10^6$ Monte Carlo steps per site. The statistical errors
are smaller than the symbol size. Jumps in the energy (see arrows)
occurred while taking the data points on the vertical lines.
}
\label{hyslp}
\end{figure}

No signs of a phase transition were seen for $p \approx 1$,
but for $p=20$ there is a clear jump in the energy as a function of $K$.
Fig. \ref{hyslp} shows the resulting hysteresis for a system
of size $L=48$. For the XY model a similar narrow hysteresis loop
was observed by DSS, but today's computers yield a clearer picture
in the Heisenberg case.

\begin{figure}
\epsfig{figure=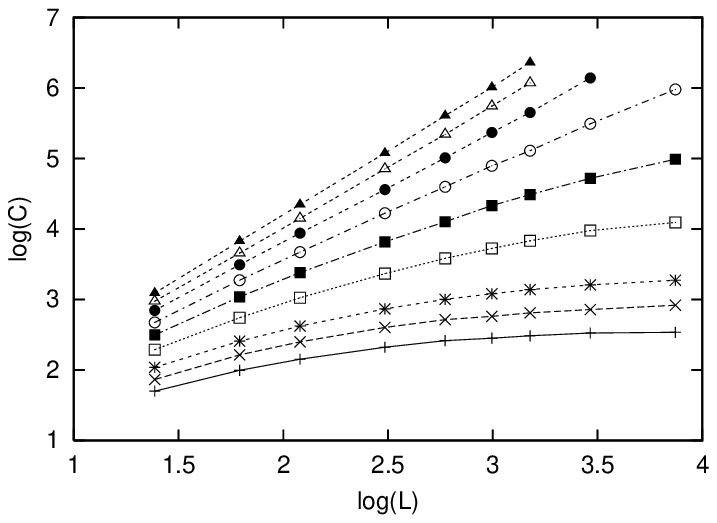}
\caption[xxx]
{Specific-heat maxima $C_{\rm max}$ 
versus system size $L$ on logarithmic scales.
The curves serve only to guide the eye. 
The data points apply to $p=6$ $(+)$, $p=7$ $(\times)$,
$p=8$ $(+\mbox{\hspace{-3.0mm}}\times)$,
$p=10$ ($\Box$), $p=12$ ({\tiny $\blacksquare$}), $p=14$ $(\circ)$, $p=16$
$(\bullet)$, $p=18$ ($\triangle$), and $p=20$ $(\blacktriangle)$.
These data suggest that the critical point at the end of the first-order
line lies near $p=16$. Each data point was determined from several
Monte Carlo runs which, because of slow relaxation, had to be long
(up to about $10^8$ updates per site each). This is
where the bulk of the computational effort went. The errors
do not exceed the symbol size.}
\label{cmaxf}
\end{figure}

Similar Monte Carlo runs  for $p<20$
show a weaker first-order character,
but do not clearly show where the first-order line ends.
In order to answer this question, we have determined the specific heat
for a grid of points in the $K$-$p$ plane. We
thus found the specific-heat maxima as a function of $K$.
Fig. \ref{cmaxf} displays these maxima $C_{\rm max}(p,L)$ versus $L$. 
In the absence of a phase transition 
$C_{\rm max}(p,L)\simeq\mbox{cst}$
when $L$ increases; this behavior is seen for small $p$. In its presence
we expect, at large $L$, 
\begin{equation}
C_{\rm max}(p,L) \simeq c_0 L^{2y-2}
\label{cmax}
\end{equation}
with $y=2$ ($y<2$) in the case of a first-order
(continuous) transition.
The data for $p=20$ in Fig.\,\ref{cmaxf} are consistent with $y=2$. The
finite-size divergence weakens for $p<20$, and the $p=16$ data indicate a
continuous transition with $y=1.84 \pm 0.05$. The downward trend at even
smaller $p$ is consistent with $C_{\rm max}(p,L)\simeq\mbox{cst}$ 
at large $L$. This suggests that the first-order line in
the $p$-$K$ diagram ends in a critical point near $p=16$. 

\begin{figure}
\epsfig{figure=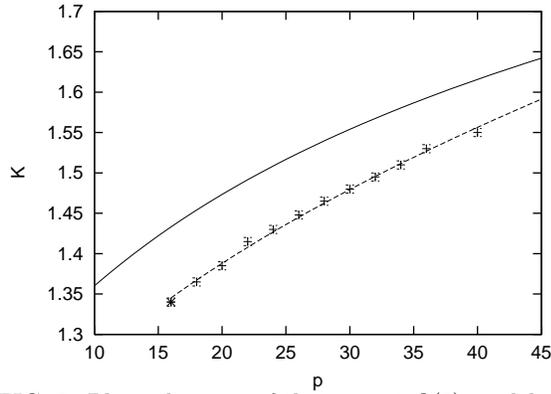}
\caption[xxx]
{Phase diagram of the present $O(3)$ model in the $p$ vs. $K$ plane. The
full curve is obtained by equating the high-$T$ and the
low-$T$ expansions of the free energy. The data points
represent our numerical results.
}
\label{htltpd}
\end{figure}

Simulations for $p>20$ show an enhanced first-order character.
Transition points were found by several runs, starting with half the
system fully aligned, and the other half chosen randomly. 
The results, which hardly depend 
on $L$ for $L>32$, are shown in Fig. \ref{htltpd} versus $p$.

The transition points can also be estimated from the 
high- and low-$T$ expansions of the free energy. Neglecting
loop diagrams in the high-$T$ expansion the lattice effectively
reduces to the Bethe lattice (BL). Its
partition function `per bond' is
\begin{equation}
\int {\rm d} \vec{s} \; \exp\;
            2K[(1+\vec{s} \cdot \vec{t})/2 ]^p\,=\, 
4 \pi \sum_{k=0}^\infty \frac{(2K)^k}{(1+pk) k!}
\end{equation}
where the prefactor accounts for the phase space volume of a spin and
the sum for the spin-spin interaction. For $N$ spins and $zN/2$ bonds we
thus have
\begin{equation}
Z_{\rm BL}  =  (4 \pi)^N
\left( \sum_{k=0}^\infty \frac{(2K)^k}{(1+pk) k!} \right)^{zN/2}
\end{equation}
which yields the high-$T$ approximation $F_{\rm HT}$ of the free energy
of a square lattice ($z=4$) of $N=L^2$ sites as
\begin{equation}
\frac{F_{\rm HT}}{Nk_{\rm B}T}  = - \log(4 \pi) - 2 \log
\left( \sum_{k=0}^\infty \frac{(2K)^k}{(1+pk) k!} \right)
\end{equation}
At low $T$, the spin-wave approximation (SWA) of ${\cal H}$ is 
\begin{equation}
{\mathcal H}_{\rm SWA}(\{\vec{s}_i\}) =
-4NK+{\mathcal H}_{\rm G}(\{{s}_i^x\})+
{\mathcal H}_{\rm G}(\{{s}_i^y\})
\end{equation}
where ${\mathcal H}_{\rm G}$ is the Gaussian Hamiltonian
\begin{equation}
{\mathcal H}_{\rm G}(\{{s}_i\}) = \frac{1}{2} p K \sum_{<ij>} (s_i-s_j)^2
\label{HG}
\end{equation}
By standard methods one obtains from it 
the low-$T$ approximation $F_{\rm LT}$ 
to the free energy,
\begin{eqnarray}
\frac{F_{\rm LT}}{Nk_{\rm B}T}  &\simeq& 
-4K - \log(4 \pi) + \log(8pK) \nonumber\\
&&+\frac{1}{N} \sum_{m,n=0}^{L-1}\!\!\!{}'\,
\log[(\sin\frac{\pi m}{L})^2+ (\sin\frac{\pi n}{L})^2]
\end{eqnarray}
where the prime indicates that $(m,n)=(0,0)$ 
is excluded from the sum. 
For large $N=L^2$ the sum on $m$ and $n$ tends towards
$-2 \log 2 + 4G/\pi = -0.2200507\cdots$ where $G$ is Catalan's constant.

The intersection of the two free-energy branches was found numerically for
several $p$. The resulting approximation of the first-order line, shown in
Fig.\,\ref{htltpd}, is in a good qualitative agreement
with the Monte Carlo results.\\

\begin{figure}
\epsfig{figure=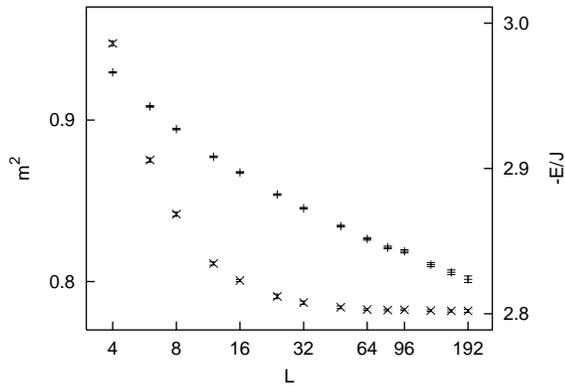}
\caption[xxx]
{Magnetization squared ($+$) and energy($\times$) per spin versus system
size $L$ in the low-$T$ phase at $K=1.4$, $p=20$.}
\label{m2evl}
\end{figure}

Next, we check the consistency of our magnetization data for the
low-$T$ phase with the Mermin-Wagner theorem \cite{MW,Mermin}.
Fig.\,\ref{m2evl} shows that the mean square magnetization
$m^2\equiv L^{-4}\sum_i\sum_j\langle\vec{s}_i\cdot\vec{s}_j\rangle$
decays slowly with $L$. 
In contrast, the energy
rapidly tends to a constant with increasing $L$. 

In order to compare this magnetization
behavior to theory, we recall that in the 
standard ($p=1$) Heisenberg model the correlation length $\xi$ is well
fitted \cite{ST}
at low $T$ by $\xi(K)\approx C\exp(2\pi K)/(1+2\pi K)$ with
$C\approx 0.01$.
For $1\ll r\lesssim\xi$ one expects the SWA result
$g(r)\equiv\langle\vec{s}_i\cdot\vec{s}_{i+r}\rangle\sim r^{-\eta}$ 
to hold, where $\eta=1/\pi K$. 
Consequently
$m^2\sim L^{-2}\int_0^L{\rm d} r\,rg(r)\sim L^{-\eta}$ for 
$1\ll L\lesssim\xi$. For $\xi\lesssim L$ the integral on $r$ converges
at the upper limit and one has $m^2\sim L^{-2}$.

Now take $p\gg1$ in the model under study.
Then the angle $\theta$ between
two neighboring spins is in a narrow two-dimensional harmonic
potential well as long as $\theta\ll \pi p^{-1/2}$.
For $\pi p^{-1/2}\lesssim\theta$
the Boltzmann weight is decreased by a
factor $\exp(-2K)$ and almost independent of $\theta$. 
When $K\gg 1$, most angles are small, and 
$g(r)$ will behave according to the SWA, but with an  
exponent $\eta=1/\pi pK$; and the correlation length
$\xi(pK)$ estimated as above will
exceed any system size $L$ attainable in simulations (disregarding
a renormalization effect of $\xi$ due to the nonlinearity of $V$). 

Next let $K\sim 1$ while still $pK\gg 1$.  Then the fraction
of nearest neighbor spins with large relative angles
will no longer be exponentially small in $K$. 
This will cause a
downward renormalization of the effective coupling of the SWA, if this
concepts remains at all applicable, and of $\xi$,
but it is not {\it a priori} clear if $\xi$
will still exceed the system size. 
To answer this question we consider Fig.\,\ref{m2evl}.
For $p=20$ and $K=1.4$
the unrenormalized SWA gives $\eta=1/\pi pK=0.012$.  
Fig.\,\ref{m2evl} confirms the power law decay of $m^2$, but
yields a renormalized exponent  
$\eta_{\rm eff}\approx 0.030$, estimated from the range 
$32\leq L\leq 192$.
This corresponds to an effective SWA
coupling $K_{\rm eff}\approx 10.6$. We note that $\xi(K_{\rm eff})$
is still very much larger than our $L$
values, which indicates the self-consistency of the
renormalized SWA.
Hence we conclude that the low-$T$ phase has a
correlation length $\xi$ much larger than the system sizes $L$
considered here, and has a pair correlation that, at these
distances, decays as a power law.\\ 

\begin{figure}
\epsfig{figure=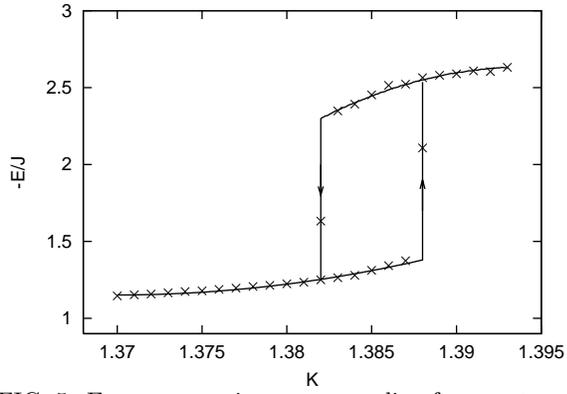}
\caption[xxx]
{Energy per spin versus coupling for a system of size $L=192$ and $p=20$
with antiperiodic boundary conditions. The first-order character is still
apparent under these conditions. Each data point represents a simulation of
$2 \times 10^5$ Monte Carlo steps per site. The statistical errors are 
comparable to the symbol size.}
\label{hysla}
\end{figure}

Our finite sizes $L$ restrict the spin waves to small deviations, so
that $m^2$ is considerable. One may ask how stable the first-order
transition is under large deviations occurring in large systems.
We have imposed large-amplitude waves using antiperiodic boundaries in
both directions.  This reduces $m^2$ considerably in finite systems
at low-$T$, and renders the low-$T$ phase less stable. Monte Carlo data
at $p=20$, $L=48$ show that the
energy jump and hysteresis are strongly suppressed. The deformation
energy per bond is $\propto L^{-2}$. Fig. \ref{hysla} shows that for
$L=192$ indeed the first-order character is partly restored to the
situation of Fig. \ref{hyslp}.
This indicates that the first-order transition persists even when
spin waves suppress the magnetization at large $L$.

It is not clear how to define an order parameter reflecting a symmetry
of the model.  The phases separated by the first-order line have different
degrees of short-range order, as is the case in a gas-liquid
system. Thus we expect the first order line to end  in an Ising-like
critical point. 
Indeed, our result $y=1.84 \pm 0.05$ agrees
well with the Ising magnetic exponent $y_h=15/8$. We note
that the energy fluctuations of this model correspond with the Ising
{\em magnetic} scaling field, because it is the energy that has a
discontinuity at the first-order line.

In conclusion, we have investigated a Heisenberg  model with
interactions that depend nonlinearly on the spin products. For strong
enough nonlinearity there appears a phase transition. This transition 
is unrelated to earlier claims \cite{PS}, which applied to the linear
case. But it does seem related to the DSS transition in the XY model
in the following way. Adding a term $\gamma \sum_k (s_k^z)^2$ in
Eq.~(\ref{ham1})  leads to crossover to the O(2) model as
$\gamma$ varies from 0 to $\infty$. In the $\gamma p$ plane we expect
a line $p=p_c(\gamma)$ (with $p_c(0)\approx 16$) above which the
transition is first order and below which it is of the KT type when
$\gamma>0$.

\acknowledgements
We are indebted to Prof. J.M.J. van Leeuwen for valuable discussions.
This research is supported in part by the Dutch FOM foundation (``Stichting
voor Fundamenteel Onderzoek der Materie'') which is financially supported by
the NWO (`Nederlandse Organisatie voor Wetenschappelijk Onderzoek'), and by
the ministry of education of the P.R. China.

\end{document}